# Detection of keratoconus Diseases using deep Learning

AKM Enzam-Ul Haque[1*], Golam Rabbany[1], Md. Siam[2]

[1*, 1, 2]Daffodil International University, Savar/Dhaka, Bangladesh

[1*]Enzam3401@diu.edu.bd
[1]rabbany.cse@diu.edu.bd
[2]siam15-3150@diu.edu.bd

**Abstract.** One of the most serious corneal disorders, keratoconus is difficult to diagnose in its early stages and can result in blindness. This illness, which often appears in the second decade of life, affects people of all sexes and races. Convolutional neural networks (CNNs), one of the deep learning approaches, have recently come to light as particularly promising tools for the accurate and timely diagnosis of keratoconus. The purpose of this study was to evaluate how well different D-CNN models identified keratoconus-related diseases. To be more precise, we compared five different CNN-based deep learning architectures (DenseNet201, InceptionV3, MobileNetV2, VGG19, Xception). In our comprehensive experimental analysis, the DenseNet201-based model performed very well in keratoconus disease identification in our extensive experimental research. This model outperformed its D-CNN equivalents, with an astounding accuracy rate of 89.14% in three crucial classes: Keratoconus, Normal, and Suspect. The results demonstrate not only the stability and robustness of the model but also its practical usefulness in real-world applications for accurate and dependable keratoconus identification. In addition, D-CNN DenseNet201 performs extraordinarily well in terms of precision, recall rates, and F1 scores in addition to accuracy. These measures validate the model's usefulness as an effective diagnostic tool by highlighting its capacity to reliably detect instances of keratoconus and to reduce false positives and negatives.

## 1    Introduction

A degenerative disorder affecting the cornea, the transparent, dome-shaped front surface of the eye is called keratoconus. The cornea gradually weakens and thins in this disorder, changing from having a normal spherical curve to a conical or bulging form. This impairs the cornea's capacity to precisely refract light, resulting in distorted and blurry vision. Even though keratoconus is not very common, it can significantly affect a person's eyesight and quality of life. Although the exact etiology of keratoconus is unknown, genetics, rubbing one's eyes, and environmental factors may all play a role in the condition's development. The illness usually starts in adolescence or early adulthood and gets worse with time.

In this phase, we assembled a dataset comprising 4011 photos related to keratoconus. This dataset included 442 eyes from 280 patients, divided into three groups: 204 eyes from 104 patients were normal eyes (NOR), 215 eyes from 113 patients had keratoconus (KCN), and 123 eyes from 63 patients were suspected of having keratoconus (SUSPECT). NOR, KCN, and SUSPECT had mean ages (± SD) of 33.4 (±10.1), 29.0 (±9.3), and 28.6 (±9.4) years, respectively. To be more precise, 58 eyes with KCN and 56 normal eyes were gathered using various Pentacam settings. 150 eyes from 85 patients made up an independent validation subgroup that was acquired from the de Olhos-CRO private hospital. 50 KCN eyes from 31 patients, 50 suspect KCN eyes from 25 patients, and 50 normal eyes from 29 participants made up this group. In the validation subset, the mean ages (± SD) of NOR, KCN, and SUSPECT were 29.5 (±4.7), 26.3 (±6.8), and 29.1 (±5.3) years, respectively.

We used a broad strategy in our first research by utilizing the strengths of many deep learning CNN architectures. We examined the nuances of DenseNet201, InceptionV3, MobileNetV2, VGG19, and Xception in particular, applying these models to a thorough examination of our dataset. We were able to take use of the unique qualities and characteristics of each design through this comprehensive investigation, which opened up new avenues for our comprehension of the information and the underlying patterns.

This study is organized as follows: an extensive review of the literature, an explanation of the experimental design, a presentation of the experiment results, a discussion, and a conclusion. The paper also examines its limitations and possible directions for further investigation.

## 2   Literature review

A wide range of subjects are covered in the literature review, such as analyses of keratoconus disease, 2D CNN networks and the investigation of knowledge gaps for the classification of keratoconus disorders using 2D CNNs.

### 2.1 keratoconus disease literature review

Currently available supervised artificial intelligence (AI) models for keratoconus (KCN) detection typically show high accuracy, with area under the receiver operating characteristic curves (AUCs) ranging from 0.90 to 1.0. These models often use Pentacam or a combination of Pentacam and optical coherence tomography (OCT) parameters. These results highlight the strong potential of AI models as useful instruments for KCN identification and diagnosis [1,2,3].

Traditional machine learning methods, like as decision trees, neural networks, and discriminant analysis, have been used in previous research to evaluate corneal topography characteristics for the purpose of keratoconus (KCN) identification [3,4,5,6]. While some models assessed keratoconus (KCN) just using anterior corneal topographic maps, others broadened their analysis to include posterior corneal data [7,8,9].

Supervised models have encountered difficulties in extrapolating findings to various keratoconus (KCN) phases and sample sizes [10, 11]. On the other hand, based on variables like topography, elevation, and pachymetry, unsupervised machine learning algorithms that are not reliant on prelabeled data have successfully recognized KCN severity levels and predicted people in need of invasive corneal surgery [12].

### 2.2 2D CNN networks
Key features include convolutional layers for feature extraction, pooling layers for dimension reduction, activation functions for non-linearity, fully connected layers for predictions, and transfer learning for efficient model development.

Figure 1 showed on DenseNet201 architecture which is a deep neural network architecture consisting of dense blocks and transition blocks.

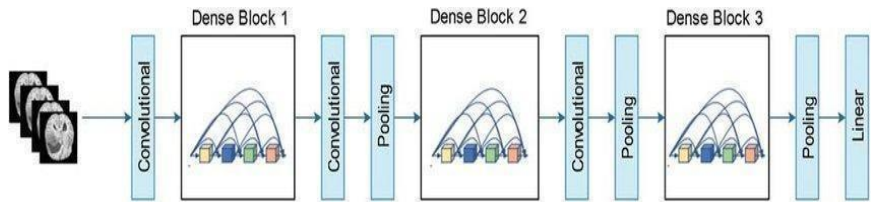

**Fig. 1.** Layered architecture of DenseNet201 [13].

InceptionV3 CNN is discussed in figure 2. Customized of Inception(GoogLeNet) developed by Google, this architecture introduced the concept of use a number of parallel filters to capture features at various scales [14].

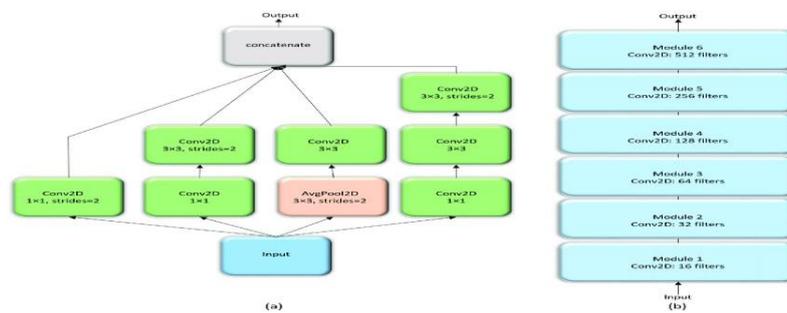

**Fig. 2.** The customized InceptionV3 CNN's appearance: A modified Inception CNN with six stacked InceptionV3 modules, and (b) the InceptionV3 module [15].

In fig. 3 MobileNetV2 structure is based on a sequence of inverted residual blocks with linear bottlenecks [16].

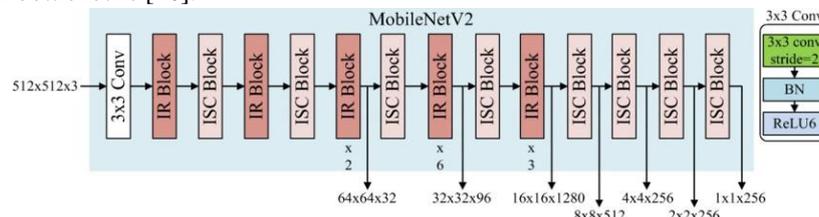

**Fig. 3.** Architecture of the MobileNetV2 backbone model [17].

Xception deep learning employs depthwise separable convolutions, which serve as a substitute for the conventional convolutions found in traditional CNNs, as depicted in Figure 4.

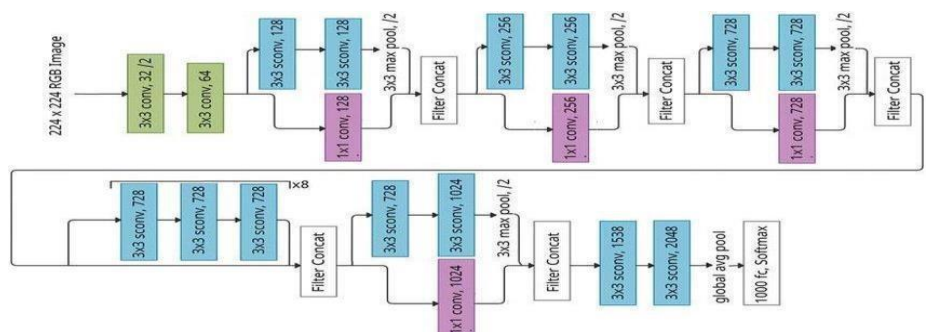

**Fig. 4.** Architecture of the Xception deep CNN model [18].

**2.3 Keratoconus disease detection Using CNN**

Using a Convolutional Neural Network (CNN) to assess corneal topography and extract information associated to keratoconus in eyes, Lavric et al. [19] proposed the KeratoDetect method. With a test dataset accuracy of 99.33%, the study showed impressive accuracy and marked a significant advancement in the use of deep learning techniques for keratoconus identification.

Tan et al. [20] achieved a 94% classification accuracy in keratoconus diagnosis using a CNN based on DenseNet121 and dynamic corneal deformation videos, showcasing notable progress in the field.

Firat et al. [21] demonstrated the efficacy of their methodology by proposing a feature vector concatenation technique that uses an SVM classifier and feature selection methods. Their methodology was able to diagnose keratoconus with an amazing 98.53% accuracy.

Using a CNN, Chen et al. [22] were able to achieve an overall accuracy of 97.85% for Pentacam HR map analysis, suggesting substantial promise for accurate keratoconus identification.

Utilizing arithmetic mean outputs from six classifiers, Kamiya et al. [23] achieved 99.10% accuracy in classifying corneal mappings for keratoconus diagnosis, highlighting the need of deep learning for extremely accurate diagnoses.

With a time-efficient framework and low computational complexity, Al-Timemy et al. [24] demonstrated a hybrid deep learning model that achieved 81.6% accuracy in keratoconus detection from corneal maps, demonstrating both precision and efficiency.

A validated model with pre-trained transfer learning networks was proposed by Al-Timemy et al. [25], achieving a remarkable breakthrough in the accurate and efficient detection of corneal diseases with keratoconus identification of 93.10%.

**Table 1.** Summary of different latest studies on keratoconus disease detection with their accuracies.

| Study | Year | CNN model | Used for detection of | Dataset/Number of maps | Accuracy Percentage |
|---|---|---|---|---|---|
| [19] | 2019 | CNN | Keratoconus | 1500 images/ 1 maps | 99.33% |
| [20] | 2022 | DenseNet121-based CNN | Keratoconus | 734 images | 94% |
| [21] | 2022 | Concatenate feature vectors, ReliefF, mRMR, Laplacian feature selection algorithms | Keratoconus | 682 images/ 1 maps | 98.53% |
| [22] | 2021 | Four color coded maps (axial, corneal thickness, front and back Elevation) | Keratoconus | 543 images/ 4 maps | 99.70% |
| [23] | 2019 | six colour-coded maps (anterior elevation, anterior curvature, posterior elevation, posterior curvature, total refractive power and pachymetry map). | Keratoconus and normal eyes | 543 images/ 4 maps | 99.10% |
| [24] | 2021 | EfficientNet-b045, hybrid DL with SVM | Normal, KCN, Suspected KCN | 692 eyes/ 7 maps | 81.6% |
| [25] | 2021 | SqueezeNet | Keratoconus and normal eyes | 2136 images/ 1 maps | 93.10% |

| | (SqN), AlexNet (AN), ShufeNet (SfN), and MobileNet-v2 (MN), LRSGD-PI classifer | | | |
|---|---|---|---|---|
| Proposed model | DenseNet201 D-CNN Model | Keratoconus, normal and suspect KCN | 4011 images/ 573 eyes/ 7 maps | 89.14 % |

**2.4 Knowledge gaps in Keratoconus disease detection Using CNN**

In Table-1, all approached methods achieve accuracy between 80% and 99%, often testing with fewer diseases for higher accuracy. Firat et al. [21] employed four color-coded maps (axial, corneal thickness, front, and back Elevation) and achieved the highest accuracy (99.70%) among all research matrix approaches. However, it's important to note that their implementation focused solely on one class, which is Keratoconus. The maximum improvement was observed in one or two classes, leading to a significant increase in accuracy. When Al-Timemy et al. [24] applied the EfficientNet-b045 hybrid DL with SVM model to three classes, they achieved an accuracy rate of only 81.6%. Achieving higher accuracy across all three classes of keratoconus disease proves to be a more challenging task and rare implementation. In our innovative exploration of keratoconus detection, we harnessed the power of the D-CNN DenseNet201 model, strategically deploying it across three distinct classes. The results spoke volumes, revealing an impressive accuracy of 89.14%. This not only signifies the effectiveness of our approach but also marks a notable leap beyond the accuracy reported by Al-Timemy et al. [24]. The success of our methodology not only contributes to the growing body of knowledge in this domain but also holds promise for advancing the precision and reliability of keratoconus diagnosis.

## 3   Research Methodology

### 3.1   Datasets

In order to improve the accuracy of the diagnosis of keratoconus disease, we assembled a dataset comprising 4011 photos related to keratoconus. The Institutional Review Board of the Federal University of São Paulo - UNIFESP/EPM granted approval as the coordinating center, and the Hospital de Olhos-CRO in Guarulhos acted as the side center. The dataset encompasses photos categorized into Keratoconus, Normal, and Suspect classes [26].

**Table 2.** The distribution of images related to keratoconus disease throughout training, testing, and validation datasets.

| Disease/Class | Number of Photos | Train Photos (70%) | Validation Photos (30%) | Test Photos (30%) |
|---|---|---|---|---|
| Keratoconus | 1400 | 980 | 420 | 420 |
| Normal | 1400 | 980 | 420 | 420 |
| Suspect | 1211 | 847 | 364 | 364 |
| **Total** | **4011** | **2807** | **1204** | **1204** |

Nevertheless, all of the raw photos were categorized into three groups that were necessary for validation, testing, and training. An example dataset with pictures of keratoconus disease is shown in Figure 5.

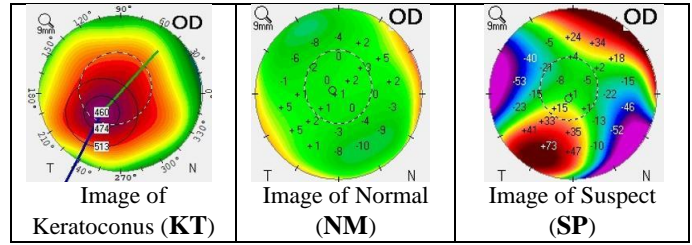

**Fig. 5.** Example of 3 classes: **KT**, **NM** and **SP**.

### 3.2 Process of experiments

We used augmentation methods in this study that included skewing, intensity modifications, vertical and horizontal flips, distortion, and random rotation and to be more precise, produced ten enhanced pictures for every original photograph. We also used an algorithm for linear contrast adjustment and a variety of filtering techniques to enhance the photos. Figure 6 offers a detailed summary of the various pre-processing techniques used in this investigation.

|  | Original Image | Centre augmentation | Centre combined augmentation | Combined augmentation |
|---|---|---|---|---|
| Keratoconus |  |  |  |  |
| Normal |  |  |  |  |
| Suspect |  |  |  |  |

**Fig. 6.** Results of image augmentation.

### 3.3 Model Training

For the CNN models, we determined the optimal hyperparameter configurations. The hyperparameters used are as follows: there were 250 epochs, a dropout rate of 0.3, a learning rate of 0.0001 and a batch size of 16. The original CNN models were trained across 250 epochs with various Early Stopping callback durations: DenseNet201 received 17 epochs of training, InceptionV3 received 16 epochs, MobileNetV2 received 39 epochs, VGG19 received 16 epochs, and Xception received 16 epochs with 10 iterations patience. The amount of training epochs that must elapse without improvement before training is discontinued is known as patience. After being optimized with the same tool, all models (DenseNet201, InceptionV3, MobileNetV2, Vgg19, Xception) were saved as .h5 files. DenseNet201 requires 31 seconds, InceptionV3-15 seconds, MobileNetV2-12 seconds, Vgg19-25 seconds and Xcption-17 seconds (s)/epoch (iterations) for model training.

### 3.4 Model Classification

At this point, we have automated the use of neural networks to identify illnesses in keratoconus. A softmax output layer was included in the training model to categorize images of keratoconus into different disease categories. Figure 7 shows the entire experimental procedure.

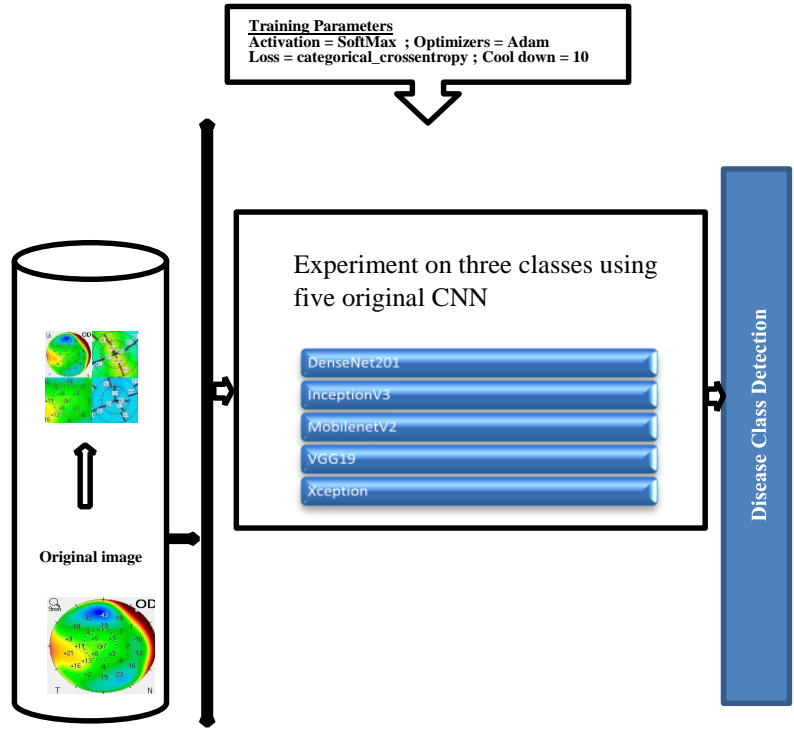

**Fig. 7.** Diagram of the experiment Models.

### 3.5    Results of experiments Experiment 1: Performance of the original CNN and their performance.

**Table 3.** Training and model accuracy of five original CNN architectures.

| Architecture | Training Accuracy (%) | Model Accuracy (%) |
|---|---|---|
| DenseNet201 | 82.51 | 89.14 |
| InceptionV3 | 81.07 | 87.13 |
| MobileNetV2 | 79.21 | 81.97 |
| Vgg19 | 81.91 | 86.89 |
| Xception | 81.33 | 87.96 |

The accuracies - as shown in Table 3, represent DenseNet201 performed best with 89.14% and MobileNetV2 had the lowest value of 81.97% **Model Accuracy**.

**Table 4.** Precision, recall, f1, and support of five (5) result of original CNN networks (based on the number of images, n= numbers)

| DenseNet201 | | | |
|---|---|---|---|
| | Keratoconus | Normal | Suspect |
| Precision | 96% | 65% | 36% |
| Recall | 98% | 36% | 46% |
| F1-score | 97% | 46% | 41% |

| Support (N) | 4210 | 419 | 363 |
|---|---|---|---|
| InceptionV3 | | | |
| Precision | 91% | 61% | 35% |
| Recall | 99% | 26% | 20% |
| F1-score | 95% | 37% | 26% |
| Support (N) | 4210 | 421 | 361 |
| **MobileNetV2** | | | |
| Precision | 99% | 48% | 26% |
| Recall | 89% | 21% | 76% |
| F1-score | 94% | 29% | 39% |
| Support (N) | 4212 | 417 | 363 |
| **Vgg19** | | | |
| Precision | 95% | 48% | 28% |
| Recall | 95% | 71% | 11% |
| F1-score | 95% | 58% | 16% |
| Support (N) | 4210 | 420 | 362 |
| **Xception** | | | |
| Precision | 99% | 56% | 33% |
| Recall | 95% | 49% | 55% |
| F1-score | 97% | 52% | 419% |
| Support (N) | 4211 | 419 | 362 |

Table 4 presents Precision, Recall, F1-score, and Specificity results for Densenet201, InceptionV3, MobileNetV2, VGG19 and Xception models across different classes. Notably, Densenet201, InceptionV3, and Xception demonstrate superior precision compared to VGG19 and mobileNetV2 on the test dataset.

KT = Keratoconus   NR = Normal   SP = Suspect

| **DenseNet201** | | | | **InceptionV3** | | | |
|---|---|---|---|---|---|---|---|
| | KT | NR | SP | | KT | NR | SP |
| KT | 4128 | 53 | 116 | KT | 4167 | 213 | 219 |
| NR | 2 | 150 | 79 | NR | 3 | 110 | 68 |
| SP | 80 | 216 | 168 | SP | 40 | 98 | 74 |

| **MobileNetV2** | | | | **Vgg19** | | | |
|---|---|---|---|---|---|---|---|
| | KT | NR | SP | | KT | NR | SP |
| KT | 3728 | 1 | 21 | KT | 3997 | 93 | 133 |
| NR | 28 | 88 | 66 | NR | 133 | 300 | 188 |
| SP | 456 | 328 | 276 | SP | 80 | 27 | 41 |

| **Xception** | | | |
|---|---|---|---|
| | KT | NR | SP |
| KT | 3989 | 4 | 27 |
| NR | 28 | 205 | 135 |
| SP | 194 | 210 | 200 |

**Fig 8.** Confusion matrix of five original CNN.

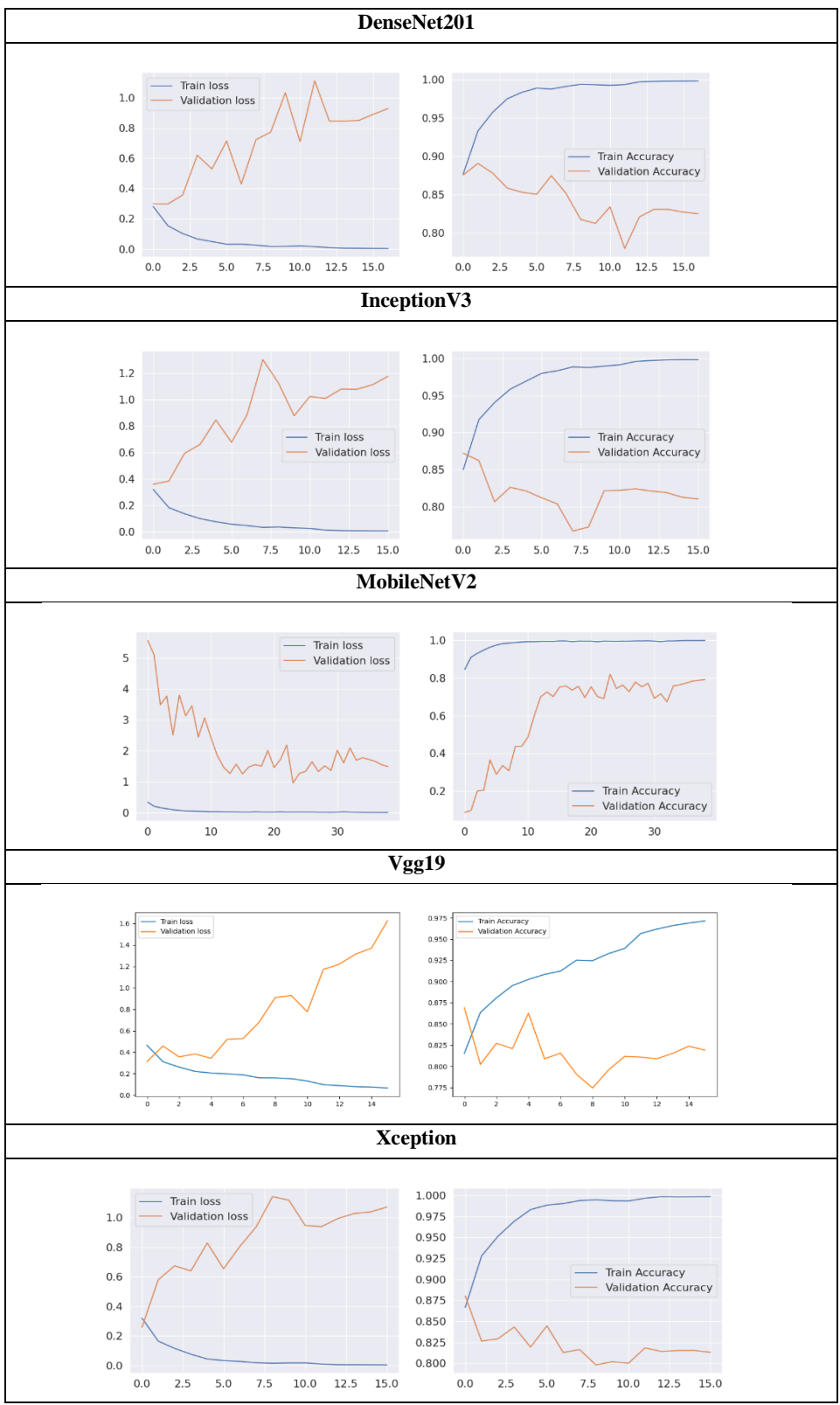

**Fig 9.** Loss and accuracy curve of five original CNN.

The five original CNNs' performance is explained by the loss and accuracy curves in Figure 9. Accuracy and loss percentages are shown on the y-axis, while the number of epochs is represented by the x-axis in the training and validation accuracy graphs of the preliminary models. There is a consistent pattern in the data across all CNN models: training and validation loss decrease with increasing epoch. As the epochs increase, the loss lines show some minor oscillations before stabilizing. The image successfully separates training and validation data, and most importantly, there is no evidence of overfitting. As demonstrated by the model's performance on training and validation data, the loss function is essential to improving the CNN architecture. This evaluation takes into account the overall amount of mistakes for every compilation and epoch. Following each optimization iteration, the loss value functions as a measure to show how successful the model is.

## 4 Discussion

We used an approaches to identify the keratoconus disease: the original CNN alone. The study we conducted was a comparison analysis in which we assessed how well five CNN-based models (DenseNet201, InceptionV3, MobileNetV2, VGG19, and Xception) classified keratoconus disorders into three different classes. The suggested model DenseNet201 demonstrated best accuracy is 89.14% over the other CNN models. With no indications of overfitting, the graphical representation shows a balanced model with appropriate segregation between training and validation data.

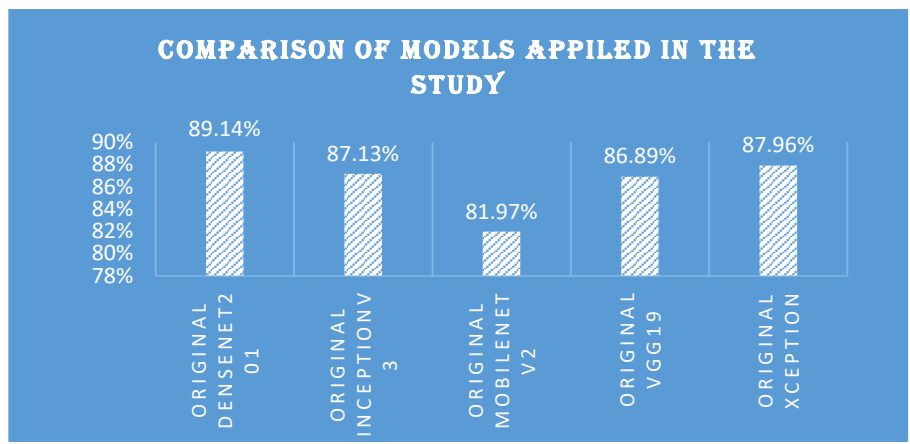

**Fig. 10.** Validation and Testing accuracy according to Image and Patch size.

## 5 Inference of the current study

Modern deep learning models are quite good at recognizing certain crops, but they are frequently ineffective at detecting diseases. CNNs can accurately detect the precise area on an eye afflicted by keratoconus disease by extracting important characteristics from photos of the condition, giving patients important information.

## 6 Future scopes

In our research, we initially implemented original D-CNN models across three crucial classes—Keratoconus, Normal, and Suspect—yielding a commendable accuracy of 89.14% in version-1. Building on this foundation, version-2 of our study aims to elevate accuracy further through the integration of an Ensemble Model (combining the strengths of D-CNN) and the computer vision transfer (ViT) model.

Recognizing the significance of achieving optimal accuracy, version-2 aligns with our goal to develop a robust application using the Python web framework FLASK. This application is envisioned to serve a diverse user base, including patients, doctors, and researchers with an interest in biomedical fields. The focus on enhancing accuracy not only fortifies the credibility of our research findings but also bolsters the reliability of the application, ensuring meaningful contributions to the medical community. As we progress to version-2, the integration of a two-class implementation further refines our approach, aligning our research with the pursuit of precision in the detection and understanding of keratoconus disease.

## 7 Limitations

The current study has challenges in obtaining a large collection of annotated photos of eye diseases, a procedure requiring knowledge from experts such as biomedical specialists. The study uses secondary data that was obtained from the public domain as opposed to primary data that was gathered in the field, which can be labor- and resource-intensive. Additionally, the testing stage is constrained by the use of free resources, such Google Colab, which provides a certain amount of server time. Future work and thought will be needed to address these issues.

## 8 Conclusion

In conclusion, we have developed a D-CNN model that shows promise for use in the biomedical industry. Our findings suggest that by accurately detecting keratoconus disorders, the advanced DenseNet201 model can increase yields in the production of eyes. This work is important for the nation, particularly in the field of biomedicine. The ultimate goal of these technologies is to simplify the diagnosis and treatment of eye conditions.